# Intrinsic Spin Transport in a Topological Insulator Thin Film

Sharadh Jois[1,a)], Gregory M. Stephen[1], Nick Blumenschein[1],

Patrick J. Taylor[2], Aubrey T. Hanbicki[1], Adam L. Friedman[1,b)]

[1] *Laboratory for Physical Sciences, 8050 Greenmead Dr., College Park, Maryland, 20740*

[2] *Army Research Laboratory, 2800 Powder Mill Rd., Adelphi, MD 20783*

## Abstract

Topological insulators (TIs) are intriguing materials for advanced computing applications based on spintronics because they can host robust spin effects. For instance, TIs have intrinsically large spin generation enabled by their large spin-orbit coupling. Furthermore, topological surface states (TSS) with spin-momentum locking and Dirac dispersion lead to long spin diffusion. Future spintronic device technology will require scalable film growth of high-quality material. We grow epitaxial films of $Bi_{1-x}Sb_xTe_{3-y}Se_y$ (BSTS, x = 0.58, y = 1) and confirm the gapless band structure with optimal doping using angle-resolved photoelectron spectra. The temperature dependence of longitudinal resistivity shows bulk transport is suppressed as temperature is decreased, and at low temperature surface transport dominates. We evaluate the spin transport properties in BSTS without using ferromagnetic tunnel contacts via a non-local resistance experiment as a function of temperature and applied charge current. As expected, these experiments reveal the necessity of decreasing the bulk conduction to best enhance the spin transport. In the TSS, we find high efficiency of charge-to-spin conversion (spin Hall angle, $\theta_{SH} \sim 1$) and spin diffusion over several microns. Further development of high-quality TIs will make them viable candidates for efficient and lossless spintronics.

## Main Text

Manipulating electronic spin as an alternate state variable for computing can potentially improve power efficiency and processor operating speed[1–3]. Magnetic random-access memory (MRAM) is the first success story of this paradigm, with multiple geometries becoming commercially ubiquitous such as stacked magnetic tunnel junctions (MTJs)[4,5], and future spin-transfer torque-MRAM variants. Advancements in next generation devices will likely include materials with high spin-orbit coupling (SOC) that can intrinsically generate spin current with a supplied charge current owing to the spin Hall effect (SHE)[6,7], offering advancements to both logic[8,9] and memory devices.

Optimization of spintronic devices depends on maximizing the charge-to-spin conversion efficiency, a metric that scales with SOC strength. For memory applications, a higher SOC strength directly relates to lower operational current, and thus higher efficiency. For logic applications, a spin channel with a high spin lifetime and a corresponding high spin diffusion length is necessary [8,10,11]. This requires combining high SOC with high mobility, which are mutually exclusive in standard materials as a high SOC usually causes significant spin scattering [12–14]. Nonetheless, the current industry pathway for spin-orbit torque memory is to incorporate conventional high-Z materials (Pt[15,16], Ta[17], W[18,19]), none of which have high enough SOC to aid improvements in efficiency.

Topological insulators[20,21] (TIs) with spin-momentum locked Dirac surface states and an insulating bulk, have uniquely high SOC (several orders of magnitude higher than industry standard heavy metals) resulting in high charge-to-spin conversion efficiency [10,11,22]. Early works on TIs using $Bi_2Se_3$ found the dominant bulk conductivity creates parallel conductance paths, limiting the utility of the novel spintronic properties of the topological surface states (TSS). Obtaining TI materials with optimal doping, where the Fermi level is in the bulk band gap, therefore became a primary challenge. Some studies demonstrate high-quality, optimally doped variations of $Bi_2Se_3$ using Sb and Te as partial substituents. Out of these efforts, balancing the composition in $Bi_{1-x}Sb_xTe_{3-y}Se_y$ (BSTS) alloys by float-zone or flux-zone growth methods emerged as a solution to create optimally doped TIs in which conduction is dominated by the Dirac TSS [23,24]. However, these materials need to be cleaved to thin layers using tape-based mechanical exfoliation methods, not amenable for large scale device fabrication. Scalable spintronic device architectures require scalable film growth, such as molecular beam epitaxy (MBE). Characterizing the spin-transport properties in these films remains an active area of interest.

In this study, we grow thin films of BSTS through MBE and probe spin generation and transport properties. The doping in these films is verified through angle resolved photoemission spectroscopy (ARPES) showing the Fermi level lying in the Dirac cone without contributions from the bulk. A non-local voltage measurement[10,25,26] and Hanle spin precession are employed to characterize the spin diffusion length ($\lambda_s$), spin relaxation time ($\tau_s$), and charge-to-spin current conversion efficiency ($\theta_{SH}$, also often called spin-Hall angle) at different temperatures and charge currents. These results show that we can efficiently generate spin currents over long distances in thin films of BSTS, further supporting the notion of TI thin films as promising candidates for low-power spin computing.

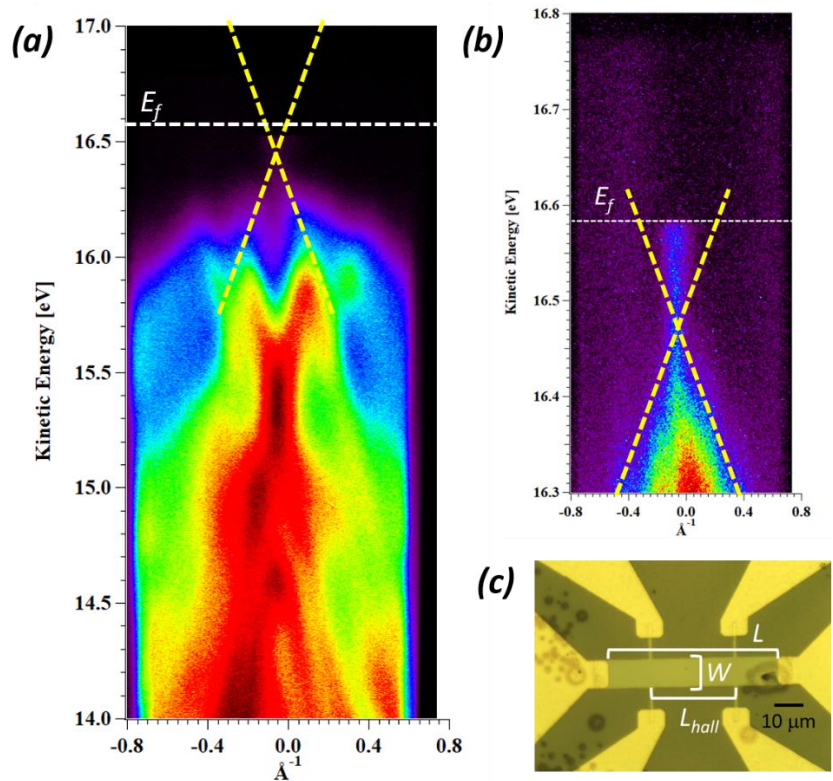

Figure 1. (a) ARPES intensity map of the BSTS film showing the bulk and surface valence bands. Finer energy resolution of within the bulk bandgap in (b) reveals a conical low-intensity band associated with the Dirac-like TSS. The dashed yellow lines are annotations to aid in visualizing the gapless Dirac cones. The horizontal dashed line is an aid to the eye to denote the position of the Fermi level ($E_f$). (c) Optical micrograph of a Hall-bar fabricated on the thin film. The dimensions of the device are width ($W = 10\ \mu m$), length ($L = 60\ \mu m$) and separation length between the Hall bar's side contacts ($L_{hall} = 30\ \mu m$).

Thin films of $Bi_{0.42}Sb_{0.58}Te_2Se_1$ were grown on InP substrates to a thickness of 12 nm, approximately two quintuple layers to reduce the contribution of bulk conduction. Additionally, the stoichiometry of the quaternary alloy was chosen to achieve an optimal doping such that the Fermi level is within the bulk band gap ($E_g \sim 0.3$ eV) and conduction is dominated by TSS. ARPES intensity maps of the pristine BSTS films shown in **Figure 1(a)** and **(b)** confirm the optimal doping in our BSTS films, consistent with previous reports [23]. We fabricated Hall bars from the BSTS film to study the transport properties, an example device is shown in **Figure 1 (c)**. The magneto-transport characterization of charge and spin in these devices is discussed below.

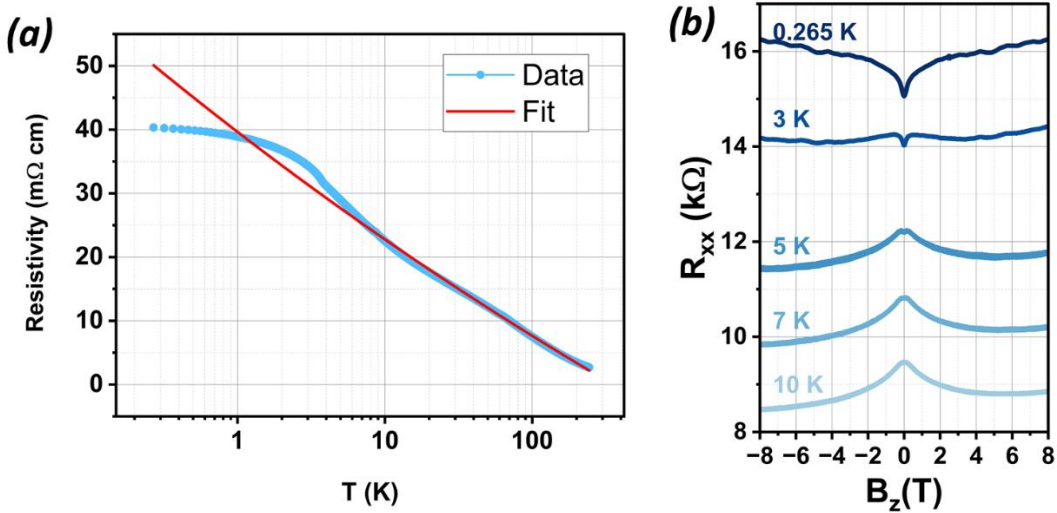

**Figure 2.** Temperature dependent transport properties in BSTS devices. (a) Longitudinal resistivity of BSTS device measured as a function of temperature (blue points) follows a power law fit (red line) $\left(\frac{164}{T^{0.47}} - 125\right)$, due to Anderson localization. Below 3 K, the resistance saturates in the purely surface transport regime. (b) Appearance of WAL cusp below 5 K is seen in the magneto resistance along with oscillations at high magnetic fields due to UCF.

The reduced bulk thickness and optimal doping evidenced by ARPES suggest minimal bulk contribution to conduction. The resistivity of the thin film devices, shown in **Figure 2 (a)**, increases with cooling and follows a power law fit down to 5 K. This suggests a disorder driven mechanism, we attribute to Anderson localization[27] effects, as well as suppression of bulk conduction, as expected. The onset of saturation in resistivity around 5 K corresponds with the surface channel dominating conduction, as also seen in previous reports on BSTS[24,28]. The longitudinal resistance ($R_{xx}$) measured with a constant current of 1 μA as a function of out-of-plane magnetic field ($B_z$), is shown in **Figure 2 (b)**. Below 5 K, these data have a cusp at low-field due to weak anti-localization (WAL). Fitting the 265 mK data with the Hikami-Larkin-Nagaoka model[21] (see *supplemental Figure S1*) yields a phase coherence length of $l_\phi = 117 \pm 5$ nm and $\alpha = -0.085 \pm 0.001$. The fit parameter α should be ½ for each conduction channel.[21] We postulate our fit of $\alpha \ll 1/2$ is due to coupling between surface states which has been seen in TI films <5 quintuple layers[29,30]. The results from magneto-transport are also consistent with the resistivity saturation seen below 5 K in **Figure 2 (a)**, where surface transport is dominant. An oscillatory signal at base temperature can be seen in the high field portion of the MR at low temperature and are magnified in *supplemental Figure S1*. These oscillations have a period of 1.5 T, and we attribute them to universal conductance fluctuations (UCF) from charge scattering[31]. The oscillations are not periodic in 1/B, ruling out Shubnikov de Haas (SdH) oscillations. The Hall effect mobility in the film at low temperature is $\mu < 10$ cm²/Vs and momentum scattering time $\tau < 10\ ps$, see *supplemental Figure S2*. These scattering parameters set limitations on the spin transport properties we discuss later on[32]. The charge transport results, combined with the ARPES, guide the spin transport studies below.

Characterizing the spin-transport of a given material usually requires additional ferromagnetic metals (FM) for injecting or detecting spins[33]. Adding tunnel barriers solves problems related to difference in conductivities[34–36] between the magnetic contacts and active material. However, tunnel barriers are plagued by pinholes and growth issues. For materials with TSSs, spin currents can be created without ferromagnetic metals or tunnel barriers due to spin-momentum locking. Thus, materials with large SOC generate transverse spin currents for a supplied charge current via the spin-Hall effect (SHE), enabling a direct measurement of the spin-to-charge conversion. The SHE is the spontaneous accumulation of opposite spins along the edges of a conductor, analogous to the classical Hall effect[6] where opposite charges are driven in opposite directions by an external magnetic field. Unlike the classical Hall effect, no external magnetic field is needed for the SHE as the SOC in the material induces the separation of spins. The SHE in high SOC materials occurs due to spin-dependent distortions of the electron trajectories[37]. Inversely, the propagation of spin current leads to accumulation of charge at the edge of a conductor due to the inverse spin Hall effect (ISHE)[7]. The non-local spin-transport experiments below utilize a combination of these effects.

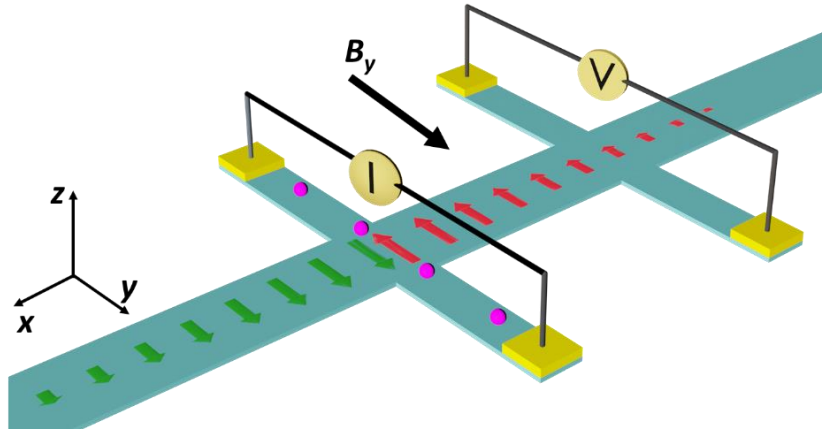

Figure 3. Schematic representation of non-local volatge measurement. When a charge current is injected between a pair of side contacts, a transverse spin current is spontaneously generated by the SHE. Due to spin-momentum locking in the TSS the spin-up (green arrow) carriers and spin-down (red arrow) carriers diffuse in opposite directions. The propagation of spin, builds up charge due to the ISHE and is measured as a non-local voltage. In-plane magnetic field is applied along the charge current direction. Together, this is referred to as the Hanle effect.

A schematic of the non-local transport configuration on a Hall bar device is shown in **Figure 3**, based on a previous theoretical proposal[26] and experimental reports on hydrogenated graphene[25], $Bi_2Te_xSe_{1-x}$ [21] and $Cd_3As_2$ [11]. A charge current is applied between a pair of contacts that are orthogonal to the channel. This causes spin diffusion in the main channel, orthogonal to the applied current. The diffusion of spins is accompanied by the accumulation of charges at the edges of the Hall bar due to the ISHE and gives rise to a measurable non-local voltage well outside the current path. Applying an external in-plane magnetic field $(B_y)$ causes the spins to precess at the Larmor frequency $(\omega_B)$ and reduces the non-local resistance $R_{nl}$ due to precessional dephasing. This Hanle effect is quantified in **Eq. 1** below. Note that conventional measurements of the Hanle effect require ferromagnetic tunnel contacts and perpendicular external magnetic field, which would destroy any TSS by breaking time reversal symmetry. Here, by employing this non-local resistance measurement at fixed current bias, we characterize spin generation and diffusion in Hall bar devices of BSTS having high SOC, without the need for FM tunnel contacts or a magnetic field that would destroy the TSS.

$$\Delta R_{nl}(B_y) = \frac{A}{\lambda_s} \times Re\left[\sqrt{1 + i\, B_y \times \boldsymbol{G}}\, \exp\left(-\frac{L_{hall}}{\lambda_s}\sqrt{1 + i\, B_y \times \boldsymbol{G}}\right)\right] \quad \textbf{Eq. 1}$$

In **Eq. 1**, the spin diffusion length, $\lambda_s$, is the dominant fitting parameter. Another important parameter is the spin diffusion time, $\tau_s$, which is embedded in $\boldsymbol{G} = \Gamma \tau_s$. Here, the gyromagnetic ratio $\Gamma = \frac{ge}{2m^*}$ is a material constant where $g = 2$ is the Lande g-factor, and $m^* = 0.32\, m_e$ is the approximate effective mass used in our fits. The effective mass value is based on previous reports on BSTS films[28]. The pre-factor $A = \frac{1}{2}\theta_{SH}^2 \rho_{sheet} W$, where $\rho_{sheet}$ is the temperature dependent sheet resistance, $W$ is the channel width and $L_{hall}$ is the separation between the non-local contacts. The spin-Hall coefficient $\theta_{SH} = \frac{J_s}{J_c}$ is a ratio of the generated spin current over the supplied charge current. The non-local resistance exhibits the first minima in the non-local signal at a resonance condition $\omega_B \tau_s \sim \frac{\lambda_s}{L}$. We measured $R_{nl}$ as a function of the in-plane magnetic field, $B_y$, at different temperatures (**Figure 4**) and current bias (**Figure 5**) to gain a complete understanding of the spin diffusion properties in BSTS. Note that a quadratic background associated with local magneto-resistance is subtracted from the raw $R_{nl}(B_y)$ data to isolate the component of $\Delta R_{nl}$ arising from the Hanle effect. Using this analytical framework, we probe spin transport in BSTS to expand on the spin generation and diffusion at different temperatures and current bias while drawing conclusions about the parallel bulk and surface conduction channels.

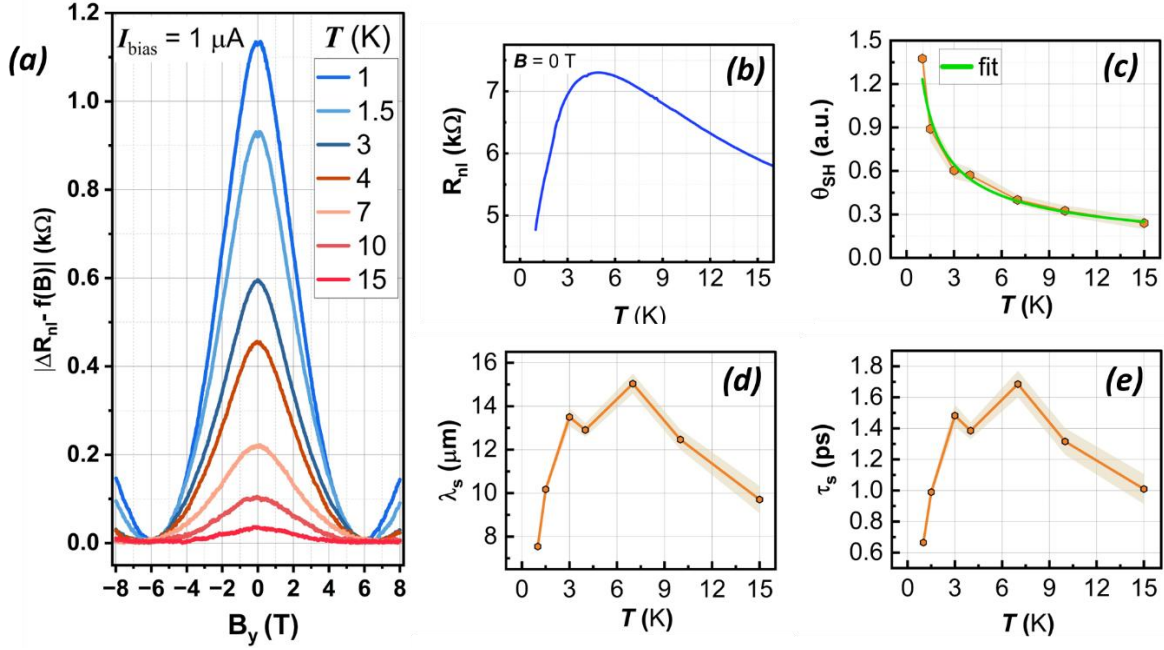

**Figure 4.** (a) $R_{nl}$ after subtraction of quadratic background at different temperatures is presented here. The spin diffusion length ($\lambda_s$) and spin diffusion ($\tau_s$) shown in panels (b) and (c). $R_{nl}$ at zero-field at 1 μA current bias measured independently shows the temperature dependence of the raw non-local signal. The spin-hall coefficient ($\theta_{SH}$) and power law fit curve in green are shown in panel (e). The inset contains the fit equation for temperature dependence of the spin-Hall coefficient and fit parameters in the table.

**Figure 4 (a)** shows the non-local resistance, $\Delta R_{nl}$, measured below 25 K at fixed current bias of 1 μA. As temperature increases, there is a clear decrease in pseudo-Lorentzian Hanle signal. The minima in $\Delta R_{nl}$ at $|B_y^*| \sim 6$ T, most prominent below 3 K, is consistent with the expected oscillatory behavior of the non-local signal at high magnetic field.[26] Using the relation $\tau_s = \frac{\hbar}{g\mu_B B_y^*}$ we estimate $\tau_s \sim 1$ ps, which is supported by fitting the data using **Eq.1**. The contribution of conventional magneto-resistance bleeding into the non-local signal, estimated using Eq. 2 [15] below, is limited by the large aspect ratio of $\frac{L_{hall}}{W} = 3$ and is estimated to be in the range $1.6 - 3.3$ Ohms.

$$R_{ohmic} = \frac{R_{sheet}}{\pi} \ln\left(\frac{\cosh\left(\frac{\pi L_{hall}}{W}\right)+1}{\cosh\left(\frac{\pi L_{hall}}{W}\right)-1}\right) \qquad \textbf{Eq.2}$$

The raw $R_{nl}$, measured independently as a function of temperature at zero-field is shown in **Figure 4 (b)**. The resistance increases as temperature is lowered, characteristic of a semiconductor, indicating bulk conduction is the dominant transport channel. Below 5 K as the TSS starts to take over the conduction channel, the resistance decreases as the temperature decreases, as expected for metallic behavior. For a more comprehensive understanding of our data, we fit the $\Delta R_{nl}$ signals using **Eq. 1** and plot the fit parameters in Figure 4 (c-e). In **Figure 4 (c)**, $\theta_{SH}$ shows a maximum of ~1.4 at low temperature and follows a power law fit $\theta_{SH}(T) \approx \frac{1.2}{T^{0.59}}$. The trend is comparable to the semiconducting bulk conduction seen in $\rho_{sheet}(T)$ which has a similar (~$T^{-0.47}$) power law dependence on temperature (**Figure 2 (a)**). The similarity between $\theta_{SH}$ and sample resistivity with temperature suggests spin generation favors the TSS channel in BSTS. Higher $\theta_{SH}$ should be achieved by reducing disorder and increasing surface transport at higher temperatures by reducing conductivity of the bulk channel. Somewhat surprisingly, the temperature dependence of $\lambda_s$ and $\tau_s$, shown in **Figure 4 (d)** and **(e)**, track with $R_{nl}(T)$ in **Figure 4 (b)**. The maximum spin diffusion appears between 3 to 8 K with $\lambda_s \sim 15$ μm and $\tau_s \sim 1.7$ ps. In this range there is a peak in $R_{nl}(T)$ so we expect this is the crossover between bulk and TSS conduction. The decrease in diffusion parameters above 7 K is likely caused by charge scattering attributed to the low mobility of materials with high SOC[22]. Below 7 K, based on $\Delta R_{nl} \propto 1/\lambda_s, \tau_s$ we would expect a divergence of these two parameters. The temperature dependence of spin transport in these epitaxial BSTS devices show that higher spin generation is linked to higher surface conduction, whereas the spin diffusion peaks at the cusp of transition from bulk to surface conduction and is limited by scattering.

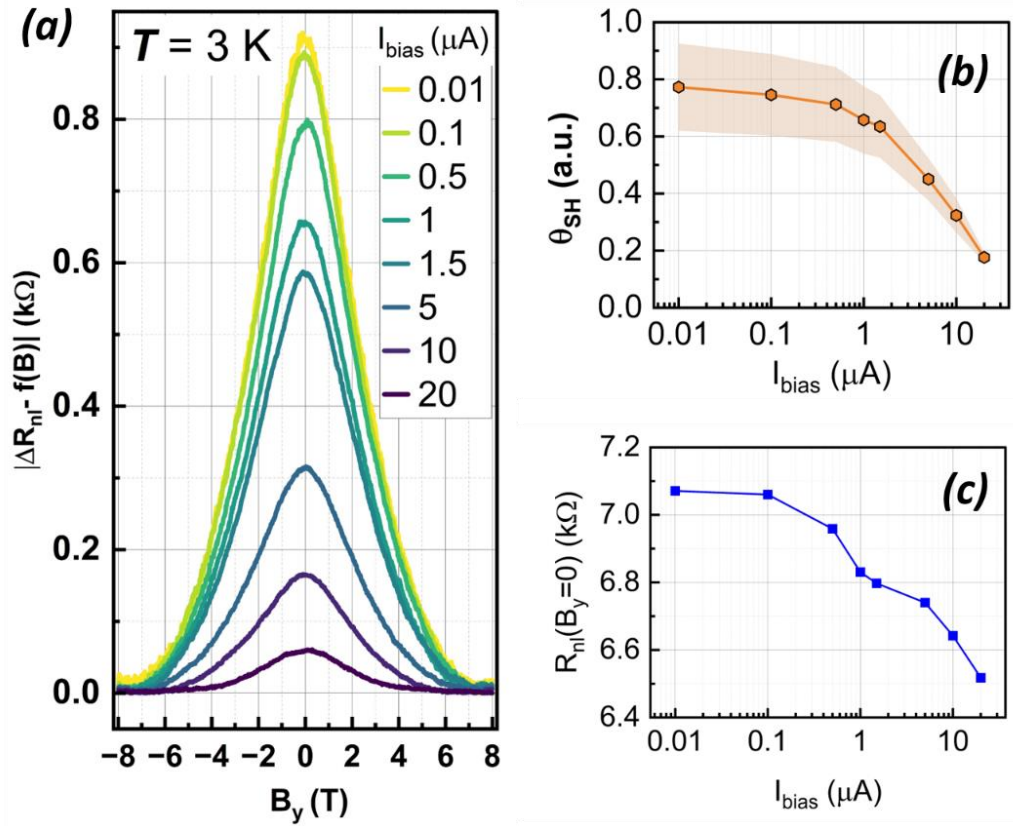

**Figure 5.** Contribution of varying current bias on the normalized non-local resistance is shown in (a) after subtraction of quadratic background. The spin-Hall coefficient θ$_{SH}$ in (b) decreases with increase in bias, similar to R$_{nl}$ in (c), following a quadratic relationship according to the solid fit curves in green and red for the respective plots. The error band in (b) appears to scale proportionally to the fit parameter.

Characterizing the current dependence of spin transport is important for low-power spintronics. To aid our understanding and complement the temperature dependence presented above, we perform similar non-local experiments and analysis at a fixed temperature while changing the current bias. The normalized non-local resistance $\Delta R_{nl}$ at 3 K, shown in figure 5 (a), decreases as current bias ($I_{bias}$) is increased from 0.01 to 20 μA. This indicates spin generation is suppressed at higher currents. This is supported by a monotonic reduction in $\theta_{SH}$ with increasing current, seen in **Figure 5 (b)**. A similar monotonic trend is seen in the fit for $R_{nl}$ at zero-field, **Figure 5 (c)**. This trend suggests there is an increase in semiconducting bulk conduction as current bias is increased. Increased scattering from disorder in the bulk conduction channel[10,11] will lead to lower $\theta_{SH}$. Increasing bias has little effect on the spin diffusion parameters as they remain around $\lambda_s = 12 \pm 1$ μm and $\tau_s = 1.2 \pm 0.2$ ps. These results indicate that despite the monotonic decrease in charge-to-spin conversion with increasing current bias, the extent of spin diffusion is largely unaffected.

In conclusion, we have characterized the intrinsic spin generation and transport properties in epitaxial films of BSTS, a TI engineered to have the Dirac TSS near the Fermi energy. Large charge-to-spin conversion $\theta_{SH} \sim 1$ is observed when conduction is dominated by surface states. Our observations also establish correlations between bulk conduction and $\theta_{SH}$ by varying temperature and applied charge current. The spin diffusion parameters in BSTS films are independent of supplied current bias, but are strongly dependent on participating conduction channels. These results provide evidence that optimally doped TI thin films are great candidates for next-generation low-power memory and logic spintronic devices.

## Supplementary Section

Additional temperature dependent electrical characterization can be found in the supplemental document.

## Author Declarations

a) sjois@lps.umd.edu
b) afriedman@lps.umd.edu

## Data Availability Statement

Data that support the findings of this study are available on request from the corresponding authors.

# *Supplementary Information*

## Results of basic Magneto-transport

The results from magneto-transport experiments at constant bias of 1 µA are given in **Figure S1**. The longitudinal resistance ($R_{xx}$) measured as a function of sweeping the out-of-plane magnetic field ($B_z$) in **Figure S1 (a)** shows conventional MR above 150 K. From 50 K to 5 K, there is a peak in $R_{xx}$ within $\pm 2$ T attributed to disorder in the film. Below 5 K, we see the appearance of oscillations due to universal conductance fluctuations (UCF) and a cusp due to weak anti-localization (WAL) effect.

The oscillatory component of $R_{xx}$ along the symmetrized positive magnetic field is shown in **Figure S1 (b)** (top panel) along with the fast Fourier transform (FFT) amplitude showing a primary period of 1.5 T (bottom panel). UCF appear due to an ensemble of Aharonov-Bohm loops in different areas of the sample and we can estimate the radius ($r$) of charge puddles, according to [1]. Using the relation $r = 0.36 \times 10^{-7} \sqrt{F} = 29.4$ nm, where $F = 0.66$ T$^{-1}$ is the frequency from FFT. This radius is used estimating the phase coherence length, $l_\phi = 2\pi r = 184.7$ nm. We explain below how this value compares with the WAL analysis.

The conductance at 265 mK is analyzed for WAL using the Hikami-Larkin-Nagaoka model [2,3], shown in **Figure S1 (c)**, based on the equations given below.

$$\frac{\Delta \sigma_{xx,\perp}}{\sigma_0} = \frac{\alpha e^2}{2\pi h} \left[ \psi\left(\frac{B_\perp}{B_z} + \frac{1}{2}\right) - \ln\left(\frac{B_\perp}{B_z}\right) \right] \text{ and } B_\perp = \frac{\hbar}{2el_\phi^2}$$

Where $\psi$ is the Digamma function, $l_\phi$ is the phase coherence length, and e and h are the electron charge and Plank constant, respectively. $\alpha$ is expected to be ~0.5 for each Dirac surface state in bulk topological insulators (TI). Our fit yields $\alpha = -0.085 \pm 0.001$ suggesting increased coupling between the surface states in TI films <5 quintuple layers in thickness, as explained in layer dependent studies [4]. The fit for the $l_\phi = 117 \pm 5$ is comparable to other reports [1,3,4] and the UCF analysis discussed above.

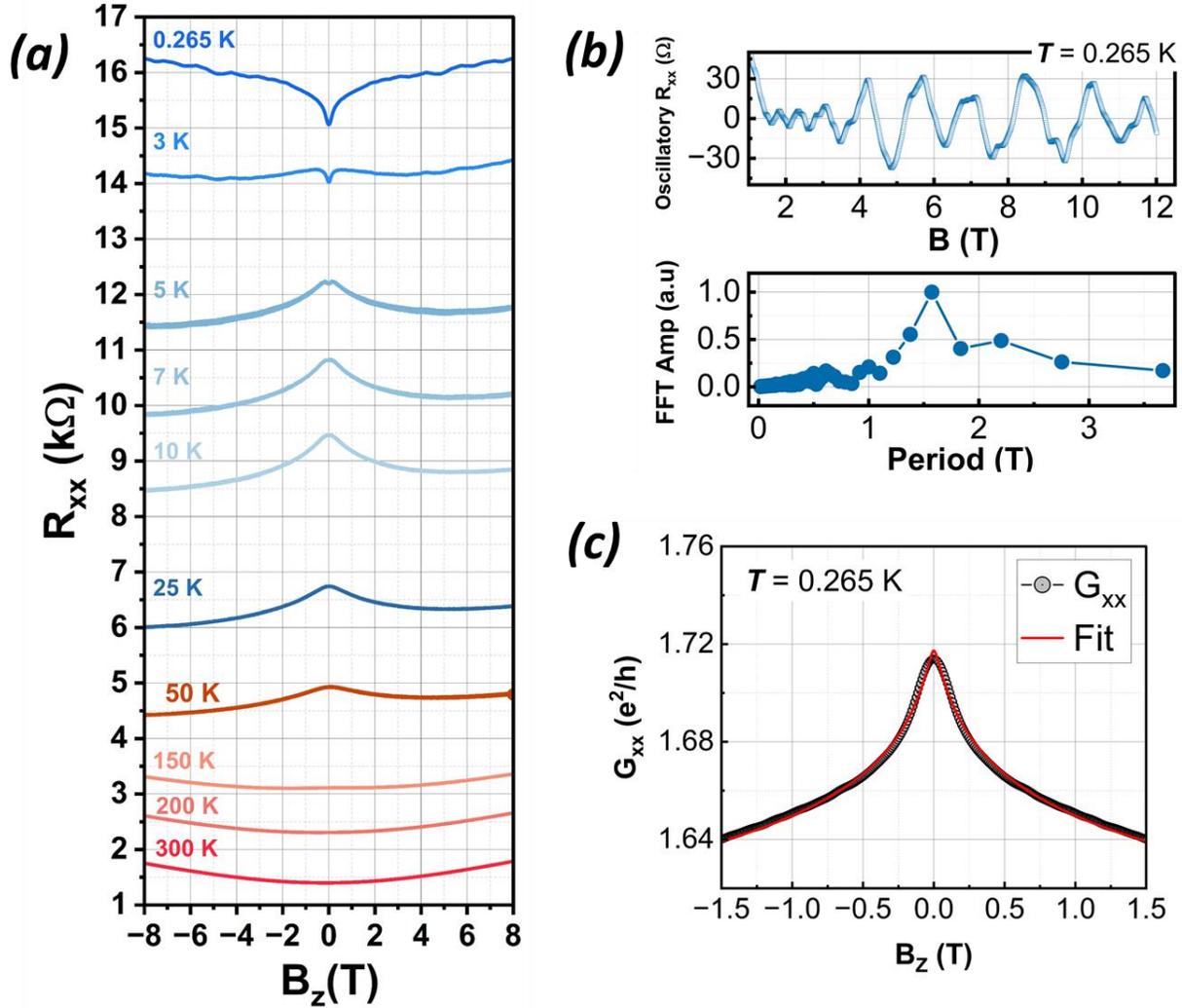

*Figure S1. (a) Magneto-resistance in BSTS at different temperatures shows the weak anti-localization (WAL) appearing below 5 K, indicating strong surface transport behavior. Parabolic conventional magneto-resistance is seen at high temperatures 150-300 K. (b) An oscillatory component in $R_{xx}$ is seen below 5 K and becomes prominent at 0.265 mK is due to UCF. The Fast Fourier Transform amplitude of this signal reveals a period of 1.5 T. (c) Based on fitting the low-temperature data using the Hikami-Larkin-Nagaoka (HLN) model, we find the phase coherence length $l_\phi = 117 \pm 5$ nm and $\alpha = -0.085 \pm 0.001$. These results are also consistent in the 3 K data.*

# Electronic Properties of the film

The transverse Hall voltage, $V_{xy}$, is measured as a function of sweeping the perpendicular magnetic field at different temperatures, with a constant current of $I = 10$ µA. The relationships for carrier density ($n$), mobility ($\mu$), and elastic scattering time ($\tau$) are given below:

$$n = \frac{IB}{e|V_{xy}|}, \qquad \mu = \frac{1}{e n \rho}, \qquad \text{and} \qquad \tau = \frac{m^*\mu}{e}$$

Here $e$ is the charge constant, $\rho$ is the resistivity of the film and $m^* = 0.32\, m_e$ is the approximate effective mass of charge carriers based on previous reports on Shubnikov-de Haas oscillations [5].

These parameters from Hall transport experiments are shown in **Figure S2 (a)**, **(b)**, and **(c)** respectively. Our observations are consistent with previous reports on TI thin films [3,4]. The carrier concentration shown in **Figure S2 (a)** ranges $2 < n < 5 \times 10^{19}\ cm^{-3}$ and is highest at room temperature when the bulk is dominant. The decrease in $n$ till 100 K is due to carrier freeze out in the bulk. The gradual increase in $n$ from 100 K to lower temperatures is due to the increasing surface conduction. The mobility and scattering time track together, reaching their maximum between 200-250 K and monotonically decreasing as temperature is reduced. The temperature dependent electronic properties discussed above confirm the dominance of the TSS in transport and further bolster the ARPES results in the main text.

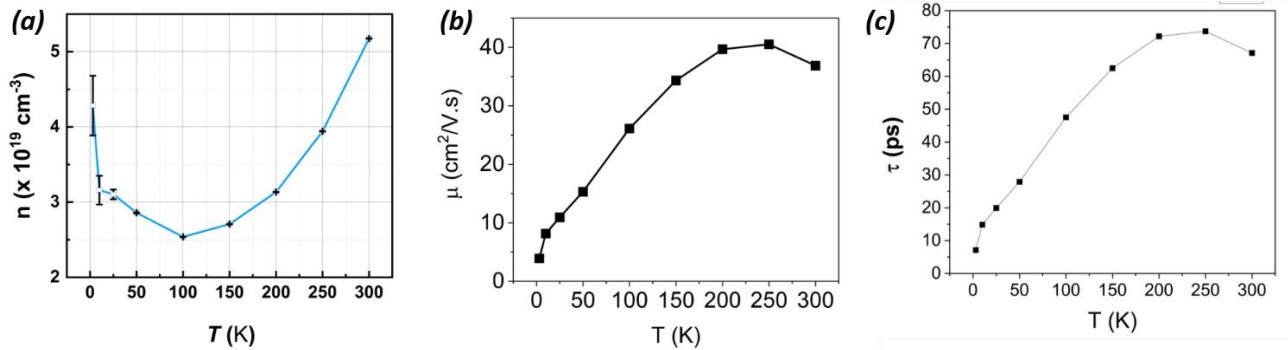

*Figure S2. (a) Carrier density, (b) mobility, and (c) scattering time as a function of temperature derived from the Hall measurements.*

# Supplemental References